\documentclass[iop]{emulateapj}

\usepackage{graphicx}
\usepackage{amssymb}

\begin{document}

\title{Temporal evolution of the Vela pulsar's pulse profile}
\author{J. L. Palfreyman, J. M. Dickey, S. P. Ellingsen,  I. R. Jones}
\affil{Department of Physical Sciences, University of Tasmania, Private Bag 37, Hobart, Tasmania 7001, Australia}

\email{jim77742@gmail.com}
\and
\author{A. W. Hotan}
\affil{CSIRO Astronomy and Space Science, 26 Dick Perry Avenue, Technology Park, Kensington WA 6151}

\begin{abstract}

The mechanisms of emission and changes in rotation frequency (`glitching') of the Vela pulsar (J0835$-$4510) are not well understood. Further insight into these mechanisms can be achieved by long-term studies of integrated pulse width, timing residuals, and bright pulse rates. We have undertaken an intensive observing campaign of Vela and collected over 6000 hours of single pulse data. The data shows that the pulse width changes with time, including marked jumps in width after micro-glitches (frequency changes).  The abundance of bright pulses also changes after some micro-glitches, but not all.  The secular changes in pulse width have three possible cyclic periods, that match with X-ray periodicities of a helical jet that are interpreted as free precession.

\end{abstract}


\section{INTRODUCTION}

The Vela pulsar (J0835$-$4510) has been the focus of many studies. Much of this has focused on either long-term timing \citep[e.g.][]{dodson2007} or short-term single pulse studies \citep[e.g.][]{johnston2001}. Here we report on both a long-term and single pulse study of the Vela pulsar.

Vela is a young, close, and bright pulsar  with characteristic age $\tau_{c}$=11.3~kyr, distance D$\approx$280~pc, flux $\mathrm{S}_{1400}$=1100~mJy, \citep{psrcat} making it a good candidate for studies using medium-sized radio telescopes. 
Vela regularly `glitches' or speeds up in rotation frequency $\nu$ with  $\Delta\nu/\nu\approx2000-3000\times10^{-9}$ \citep{rad1969,reichley1969}.

\citet{dodson2007} have previously demonstrated that Vela glitches approximately every three years. Micro-glitches, as defined by \citet{cordes1988} as $\Delta\nu/\nu\lesssim1000\times10^{-9}$ also typically occur a number of times per year \citep{alessandro1995}.

Giant pulses are considered as ones which have mean flux densities of at least 10 times the average pulse. \citet{johnston2001} reported `giant micro-pulses' and \citet{palfreyman2011} showed that Vela has bright pulses (5 times the average pulse), and consecutive bright pulses. No genuine giant pulses have been mentioned in the literature.

Vela's integrated pulse profile is well known to change with frequency \citep{cordes1978}, but despite the variation in the brightness of individual pulses there has been minimal mention in the literature of pulse profile changes in Vela over time at a fixed frequency. The exception is \citet{cordes1993} where a single tantalising sentence mentions a pulse shape variability in the Vela profile with a period of 100 days. 

In this paper we show that not only does Vela's pulse width change over time, it changes sharply after a micro-glitch, and that the rate of bright pulse activity also changes with micro-glitches. We also show that the profile changes have three quasi-periodic components (one of which has a period of $\approx100$ days) that match with periods found in X-Ray observations of a helical jet by \citet{durant2013}.

\section{OBSERVATIONS AND DATA REDUCTION}
In March~2014 we commenced a long-term single-pulse study of the Vela pulsar, collecting up to 19~hr of data each day. After 18~months we have collected over 6000~hr of single-pulse data.

We used the Mt Pleasant 26~m radio telescope (see Table~\ref{table:dish}) with a centre frequency of 1376~MHz and a bandwidth of 64~MHz. The receiver consists of a 20~cm prime-focus feed-horn with cooled dual linear polarisation feeds that are sampled with 2~bit precision at 128~million samples per second.

\begin{deluxetable}{llc}
\tablecaption{Receiving system, sampling, and data collected}
\tablehead{Component & Parameter & Value}

\startdata
Telescope & Diametre (m)					&	26			\\ \\
Location & WGS84 \\

& Longitude & 147\degr 26\arcmin 25\farcs87 E \\
& Latitude & 42\degr 48\arcmin 12\farcs90 S\\
& Altitude (m) & 65.09\\ \\
Receiver & Frequency (MHz)				&	1376		\\
& Bandwidth (MHz)				&	64			\\
& Feeds				&	Dual Cooled \\
& Polarisation & Linear		\\
& SEFD (Jy)						&	500			\\ \\
Sampling & System						&	LBA DAS		\\
& Bits				&	2			\\
& Rate (MS s$^{-1}$)	&	128		\\
& File size (s) & 10 \\ \\
Processing & Frequency channels & 16 \\
& Phase bins 					&	8192	\\	
& Resolution ($\mu$s)			&	10.9		\\ \\
Data Collected & Hours & 6000 \\
& Single pulses & $237\times10^6$ \\
& Disk usage (PB) & 1.5 \\
\enddata
\label{table:dish}
\end{deluxetable}

Each 10~s `raw' baseband file is 640~MB in size producing 4~TB of data each day. The data is then folded and phase-coherently de-dispersed using \small DSPSR \normalsize \citep{vanstraten2011} with 16 frequency channels and 8192 pulse phase bins giving a time resolution of 10.9 $\mu s$. Currently, raw files containing `interesting' events are retained, with the remainder being discarded after about 6~months. All the data is stored on the local \small RDSI \normalsize 2.3~PB storage facility for later analysis using \small PSRCHIVE \normalsize \citep{hotan2004}.

Then each file was folded in frequency, polarisation, and time, then matched to a fixed standard profile template to produce a time of arrival. Residuals for the day were then calculated using \small TEMPO2 \normalsize \citep{hobbs2006} and bad timing data were removed - typically caused by radio frequency interference (RFI) or wind stows. A best fit of $\nu$ and $\dot{\nu}$ was performed producing a residual plot appearing as white noise with a typical RMS of $\approx50\mu s$. Finally all of the day's 10~s files were phase aligned and integrated using the $\nu$ and $\dot{\nu}$ just calculated. A successful full day's observing of 19~hr would produce an integrated profile from over $7.5\times10^{5}$ individual pulses and have a signal-to-noise ratio of around $2\times10^{4}$. Days that were shortened ($\lesssim3$~hr) due to other observers or wind stows were removed from the analysis.

The arrival time residuals of these daily integrated profiles were then used to search for micro-glitches. The \small PSRCHIVE \normalsize utility `pdv' was used to produce the pulse widths at 10 and 50 per cent of the pulse heights of the integrated profile. 

We developed an algorithm to select a pulse as being `bright', but yet not include any RFI that passed through the timing residual stage, if two conditions were met:

\begin{enumerate}
\item the pulse has a peak flux of 10 or more times the average pulse.

\item the pulse at the 50 per cent level has a width (FWHM) of between $\approx160$ and $550~\mu$s (15-50 timing bins). 
\end{enumerate}

To confirm the algorithm's effectiveness we manually checked all 41442 bright pulses to verify how many were actually RFI. Only $\approx5$ per cent were falsely selected by the algorithm and these were removed from the analysis.

\section{PULSE SHAPE CHANGES}

\begin{figure*}
\center
\includegraphics[width=160mm,height=225mm]{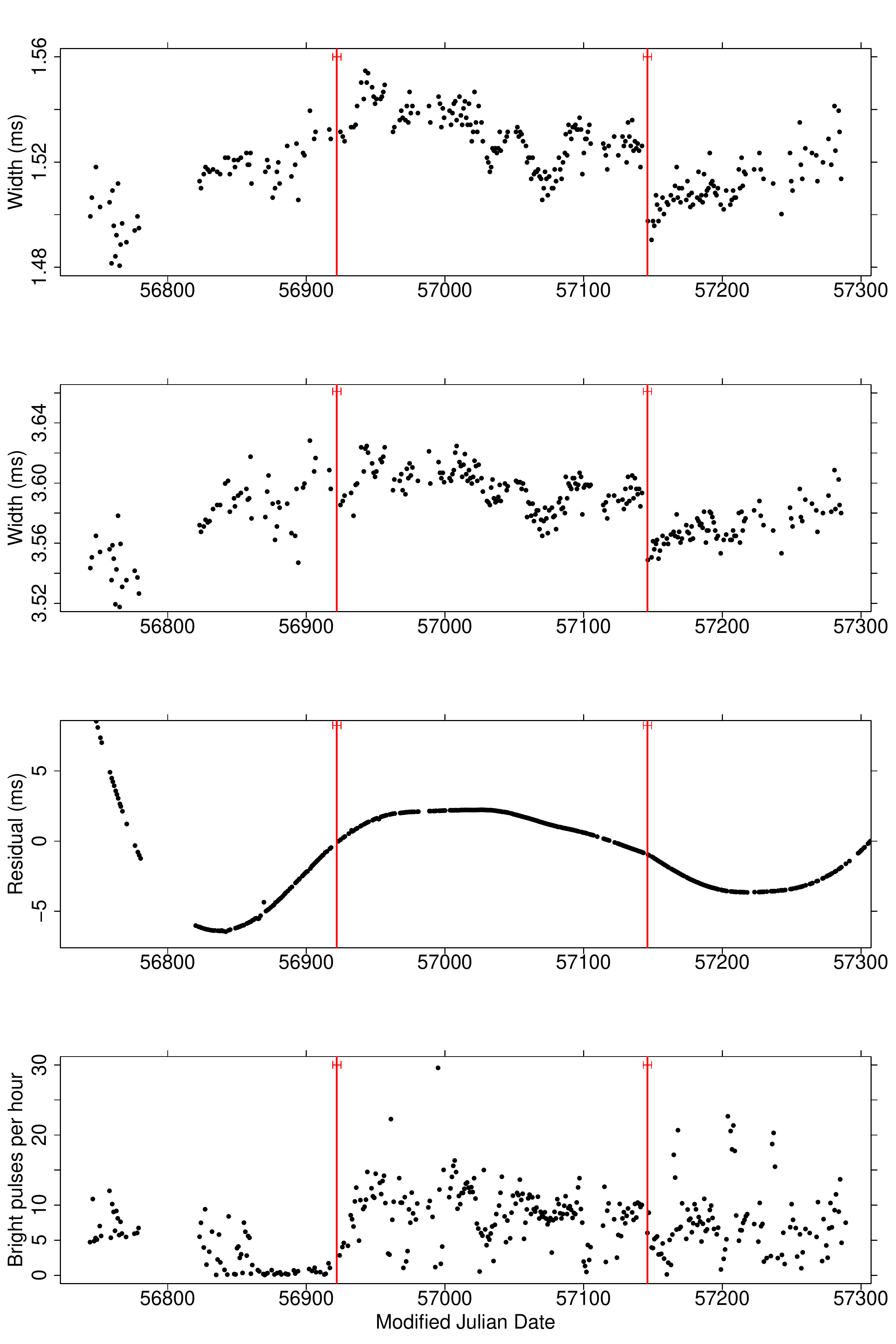}
 \llap{\parbox[b]{48mm}{\small See Figure~\ref{fig:glitchlarge} \newline \huge$\square$\\\rule{0ex}{79.5mm}}}
\llap{\parbox[b]{104.5mm}{\huge$\square$\small \newline See Figure~\ref{fig:glitchsmall}\\\rule{0ex}{77mm}}}
\llap{\parbox[b]{170mm}{\huge a\\\rule{0ex}{200mm}}}
\llap{\parbox[b]{171mm}{\huge b\\\rule{0ex}{140mm}}}
\llap{\parbox[b]{172mm}{\huge c\\\rule{0ex}{83mm}}}
\llap{\parbox[b]{173mm}{\huge d\\\rule{0ex}{28mm}}}

 \caption{(a) Plot of pulse width (in ms) at 50 per cent of the peak of the integrated pulse, (b) pulse width at 10 per cent of the peak, (c) timing residuals, and (d) bright pulse rate. Note the sudden decrease in both pulse widths when the magnitude $\Delta\nu/\nu=75.6\times10^{-9}$  micro-glitch at MJD=57143 occurred. Note the 50 day quiet period and then sudden increase in bright pulse rate after the first glitch of magnitude $\Delta\nu/\nu=0.4\times10^{-9}$ that occurred at MJD=56922. Each dot is a day of observing and all errors bars are smaller than the dots. 
}
\label{fig:width}
\end{figure*}

\begin{figure}
\center
\includegraphics[width=80mm]{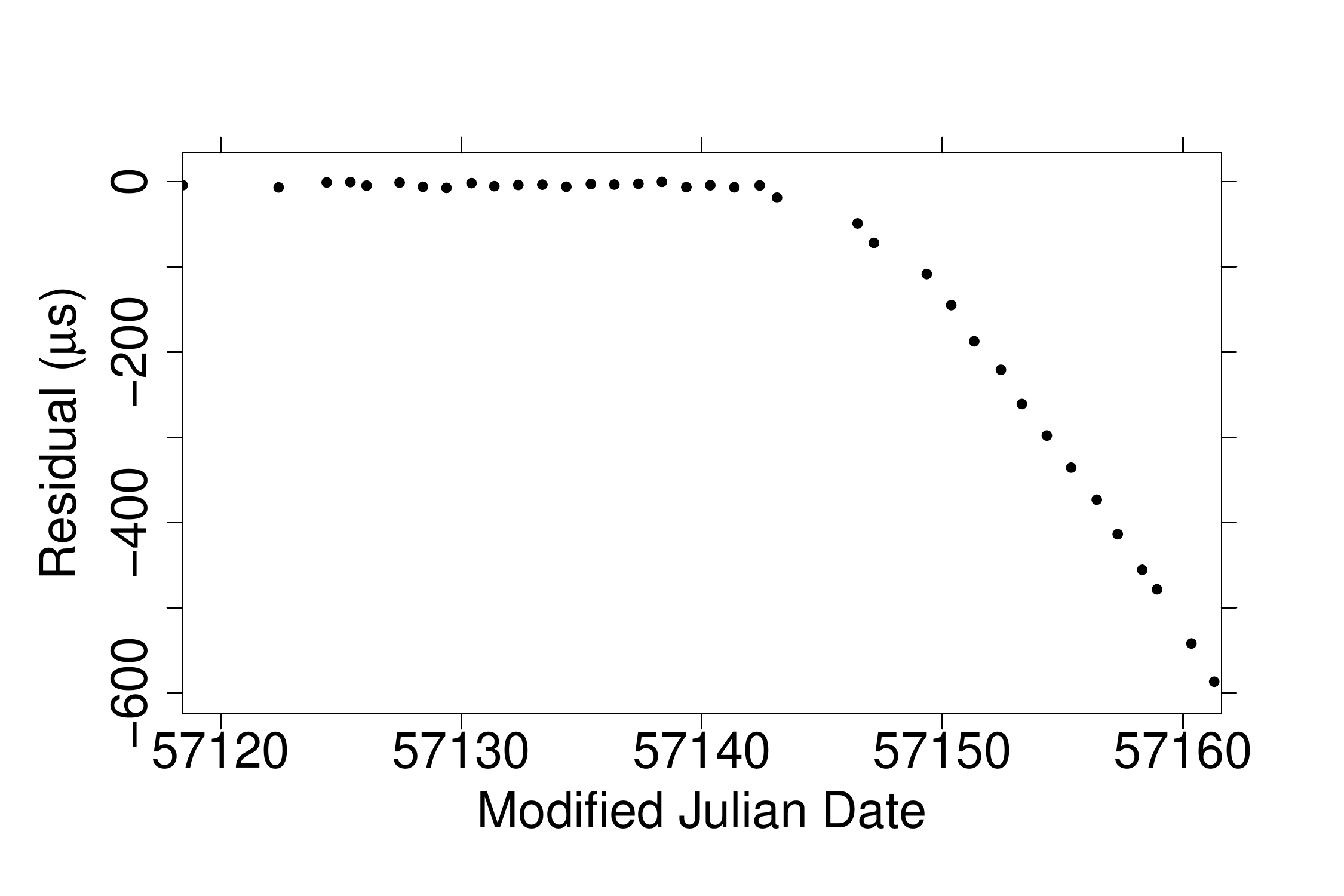}
 \caption{Daily timing residuals showing the $\Delta\nu/\nu=75.6\times10^{-9}$ magnitude micro-glitch at MJD=57143$\pm{3}$.}
 \label{fig:glitchlarge}
\end{figure}

Figure~\ref{fig:width} shows a plot of the 50 and 10 per cent profile widths, timing residuals, and bright pulse rate for MJD 56746 to 57281.
At MJD=57143 a micro-glitch of magnitude $\Delta\nu/\nu=75.6\times10^{-9}$ occurred (see Figure \ref{fig:glitchlarge}), and a noticeable and sustained reduction in the pulse width at both 50 and 10 per cent of the pulse height can be seen. 

\begin{figure}
\center
\includegraphics[width=80mm]{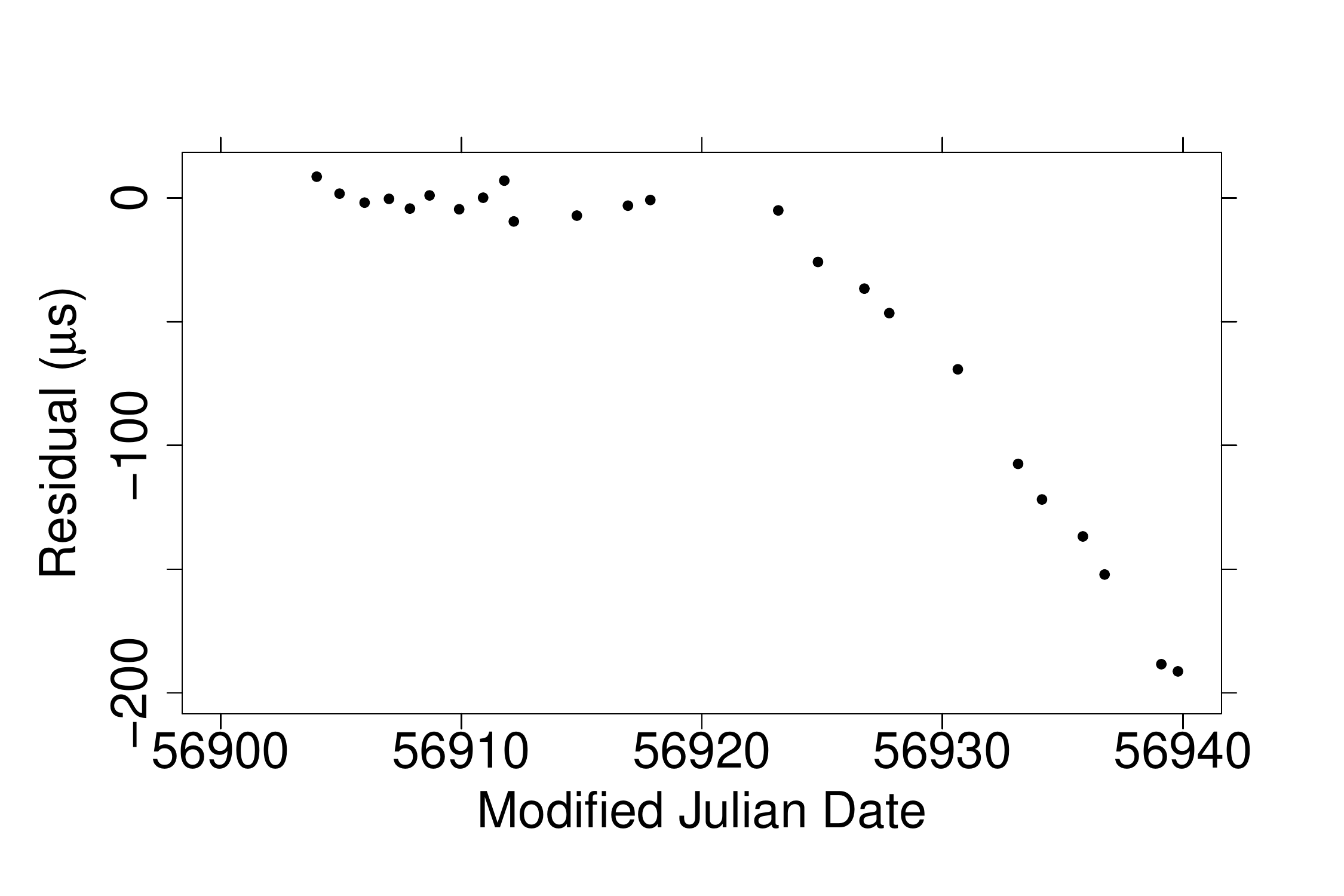}
 \caption{Daily timing residuals showing the $\Delta\nu/\nu=0.4\times10^{-9}$ magnitude micro-glitch at MJD=56922$\pm{3}$.}
 \label{fig:glitchsmall}
\end{figure}

\citet{jankowski2015}  reported a micro-glitch at MJD=56922 with a magnitude of $\Delta\nu/\nu=0.4\times10^{-9}$ (see Figure~\ref{fig:glitchsmall}). This coincided with a large increase in bright pulse activity.

\begin{figure}
\center
\includegraphics[width=80mm]{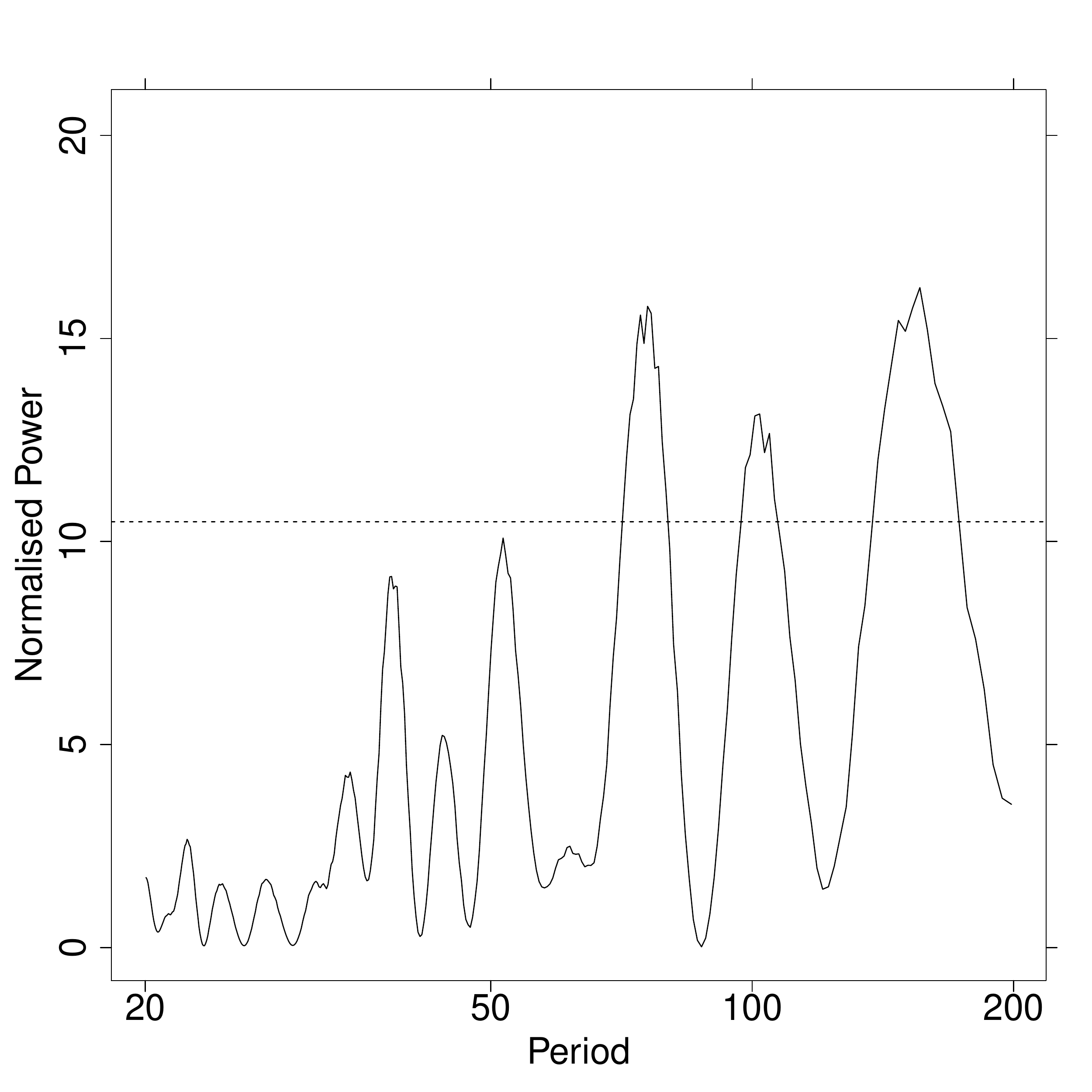}
 \caption{Lomb-Scargle periodogram of the pulse width at the 50 per cent level with a cutoff level set at p=0.001. Significant peaks are at periods of 78$\pm$5, 100$\pm$7, and 137$\pm12$ days. Due to the discontinuity, data after the micro glitch at MJD=57143 was excluded from this analysis.}
 \label{fig:lsp_w50}
\end{figure}

The plot of pulse width over a year shows distinct cyclical variations on the order of 20-30~$\mu$s. Figure~\ref{fig:lsp_w50} shows a Lomb-Scargle periodogram \citep{lomb1976,scargle1982} which has significant (p=0.001) period peaks at 78$\pm$5, 100$\pm$7, and 137$\pm12$ days. 

\begin{figure}
\center
\includegraphics[width=80mm]{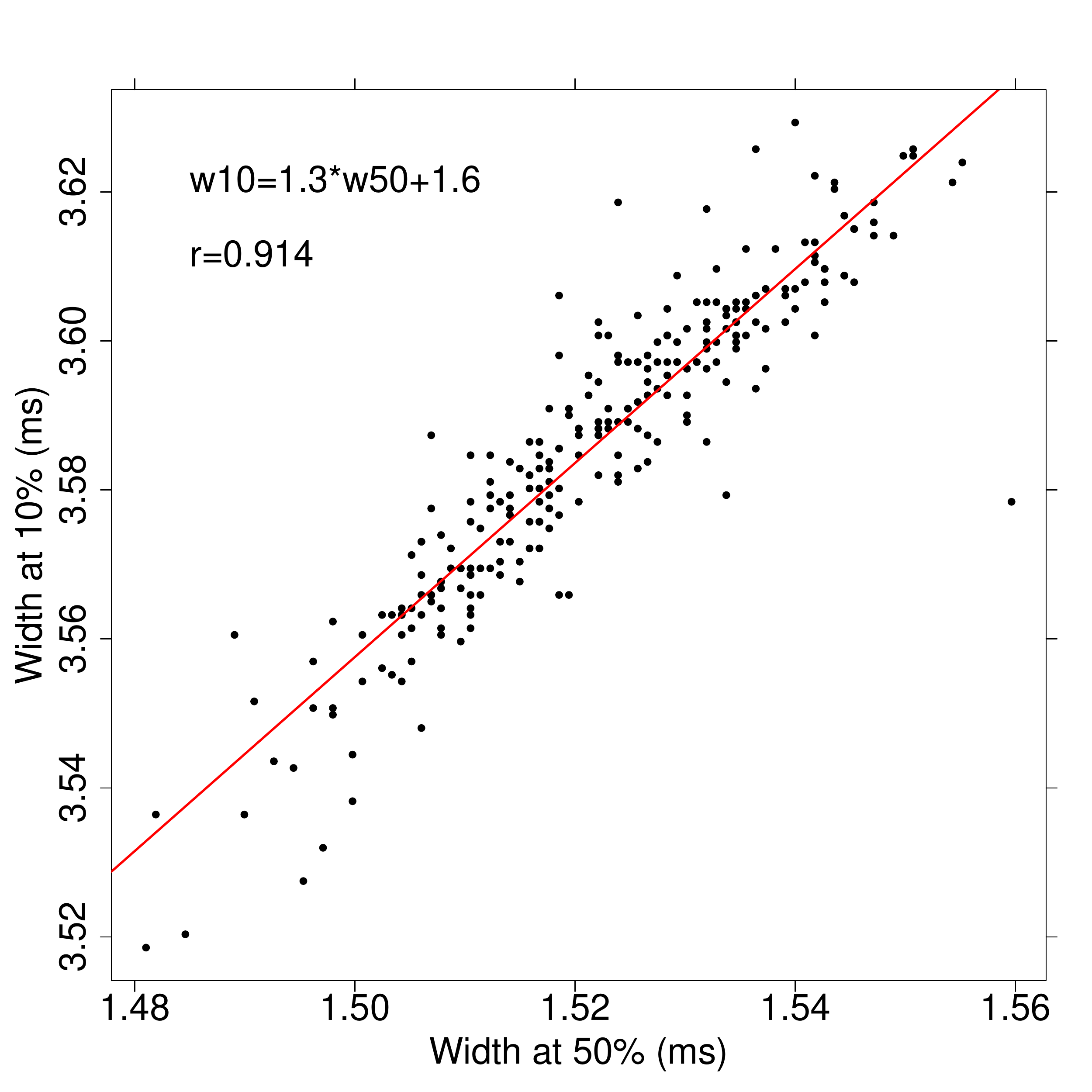}
 \caption{Correlation of pulse width at 10 per cent of the peak to 50 per cent of the peak. Low signal-to-noise days have been removed.}
 \label{fig:corr}
 \end{figure}

We also found a strong correlation (r=0.914) between the pulse widths at 10 and 50 per cent of the peak (see Figure~\ref{fig:corr}) through the entire dataset. 

This implies that what is affecting pulse width is affecting the entire pulse shape.

Our observations show that after the second and larger micro-glitch the pulse has decreased in width by about 40~$\mu$s (2.6 per cent) at the 50 per cent level and about 60 $\mu$s (1.7 per cent) at the 10 per cent level. 

Note that even though we observed with a bin width of 10.9~$\mu$s, we have from \citet{lorimerkramer2004}:
\begin{equation}
\sigma_{T}\approx\frac{W}{S/N}=\frac{5~ms}{20000}=0.250~\mu s     ,
\end{equation}

\noindent where $\sigma_{T}$ is the timing error, W is the pulse width, and S/N is the signal to noise ratio. This shows that the error bars in Figure~\ref{fig:width}a, Figure~\ref{fig:width}b, and Figure~\ref{fig:width}c would be smaller than the plotted points after integrating over 19 hours.

\begin{figure}
\center
\includegraphics[width=80mm]{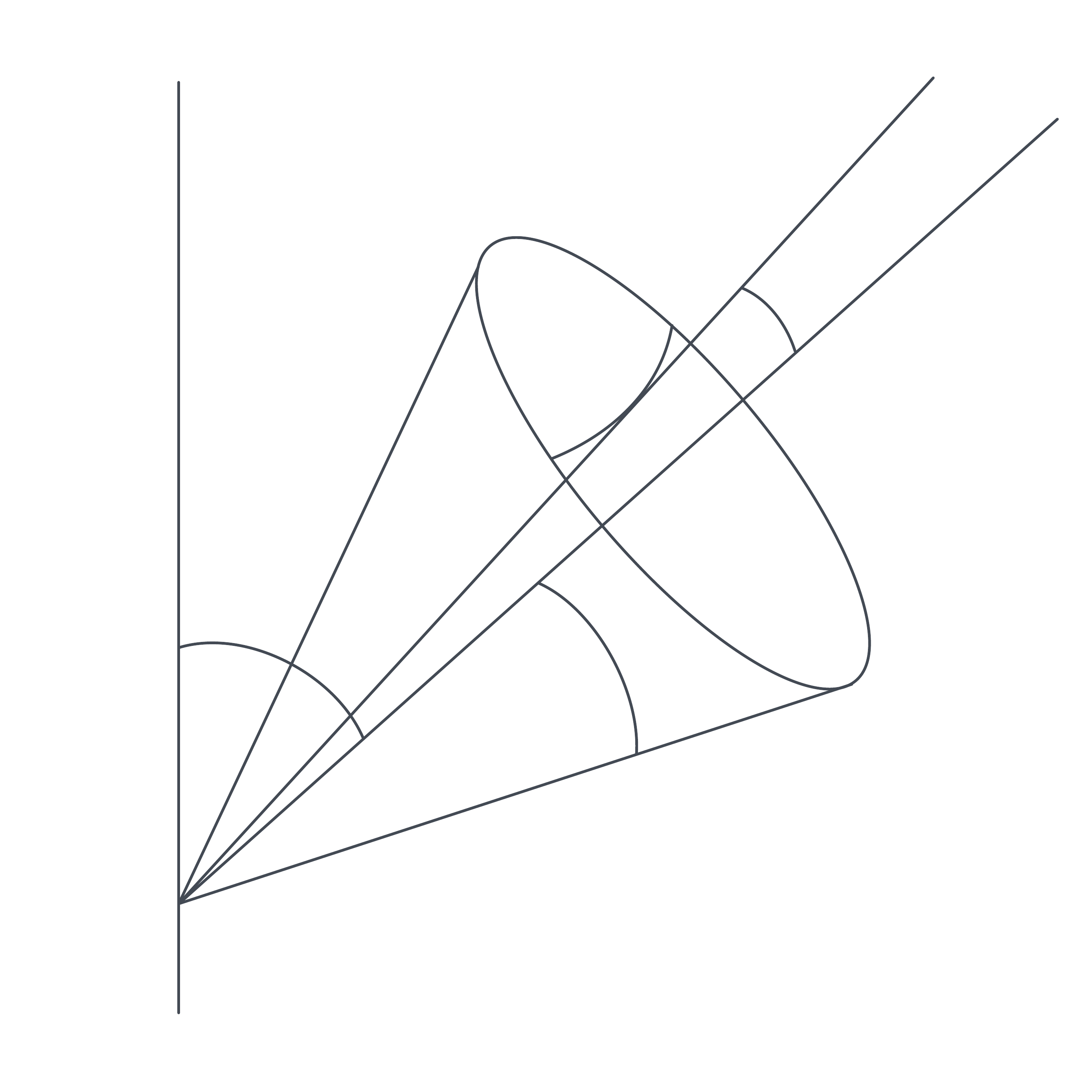}
 \llap{\parbox[b]{23mm}{$\beta$\\\rule{0ex}{57mm}}}
 \llap{\parbox[b]{36mm}{$\rho$\\\rule{0ex}{32mm}}}
 \llap{\parbox[b]{65mm}{$\alpha$\\\rule{0ex}{34mm}}}
 \llap{\parbox[b]{20mm}{To Earth\\\rule{0ex}{77mm}}}

 \caption{Simple graphic showing angles for the Vela pulsar where $\alpha$ is the angle between the rotational and magnetic axes ($55^{\circ}$), $\beta$ is the angle between the sight-line and the magnetic axis ($-6^{\circ}$), $\rho$ is the angular half width of the emission cone ($12^{\circ}$). See text for references.}
 \label{fig:pulsardraw}
\end{figure}

Pulse profile width changes like this could be caused by a slight change in our sight-line to the pulsar's magnetic axis $\beta$ (see Figure~\ref{fig:pulsardraw}) both generally throughout the year and also a step change associated with the micro-glitch. For this to occur either a change in the pulsar's rotational axis relative to the earth (geodetic precession), or a change in the magnetic axis angle ($\alpha$) relative to the rotational axis (free precession) has occurred.

\begin{figure}
\center
\includegraphics[width=80mm,height=40mm]{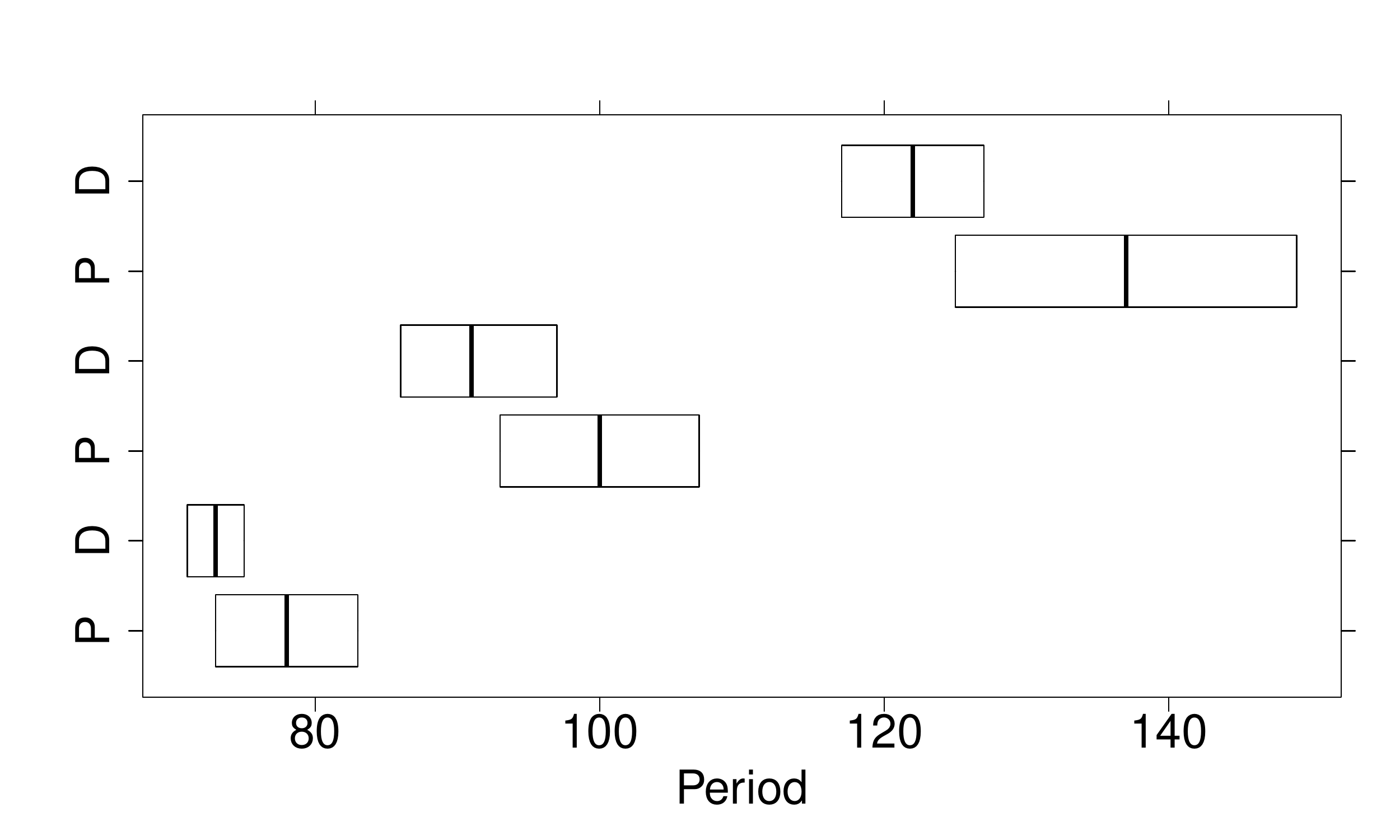}
 \caption{Periods (in days) from (P) our Lomb-Scargle plot (Figure~\ref{fig:lsp_w50}) compared to the periods from the (D) X-Ray observations from \citet{durant2013}.}
 \label{fig:periods}
\end{figure}

Precession of the Vela pulsar has been raised in the literature previously. \citet{durant2013} reveal a helical X-ray jet streaming from the rotational axis of the Vela pulsar potentially caused by precession. The jet has `acceptable' periods of 122~$\pm$~5, 73~$\pm$~2 or 91~$\pm$5 days. We see three definite periods in our pulse width data and the ranges of these fall within the ranges of the `acceptable' periods mentioned in \citeauthor{durant2013} as shown by Figure~\ref{fig:periods}. 

\citet{stairs2000} also discuss precession in a different pulsar (J1830-1059) and the periodicity in their timing residuals is very clear. However J1830-1059 does not glitch like Vela and so the cycles are easy to measure over a long period of time. They conclude that precession is the simplest explanation for their observations.

The change in $\alpha$ for Vela has also been discussed by \citet{link1997}. They contend that its very low braking index of 1.4~$\pm$~0.2 is caused by an increase of $\alpha$ over time. For Vela, $\alpha=55^{\circ}$ and $\beta=-6^{\circ}$ \citep{johnston2001} and so since $\beta$ is negative we have an inside sight-line  (see Figure~\ref{fig:pulsardraw}). An increase in $\alpha$ implies the pulse width should be shrinking overall. Our data shows a pattern of pulse width increase and then decrease following the small micro-glitch at MJD=56922. After the much larger micro-glitch at MJD=57143, we see a sharp decrease in pulse width followed by a steady increase. 

\citet{link1997} also state that the evolution of $\alpha$ is not monotonic and that it seems to be different with older pulsars as compared to younger ($t_{age}~\lesssim~10^{4}~yr$) ones.

Profile width changes could also be caused by a change in the width of the emission cone ($\rho$). This has been discussed in \citet{rankin1993} which shows a link between conal width and pulse period --- the slower the pulsar, the narrower the emission cone. Now since a glitch is a fractional increase in rotation frequency, this would imply a widening of the emission cone. With the larger micro-glitch, we observed a sharp decrease followed by a slow increase in pulse width. Regardless, the rotational increase is so small that Rankin's relation ($W=5.8^{o}P^{-1/2}$) predicts a pulse width change of only 0.178~ns. Of interest is that changes in emission cone width are linked to rotational frequency, and that a micro-glitch could be linked to a change in pulse width.

\begin{figure}
\center
\includegraphics[width=79mm]{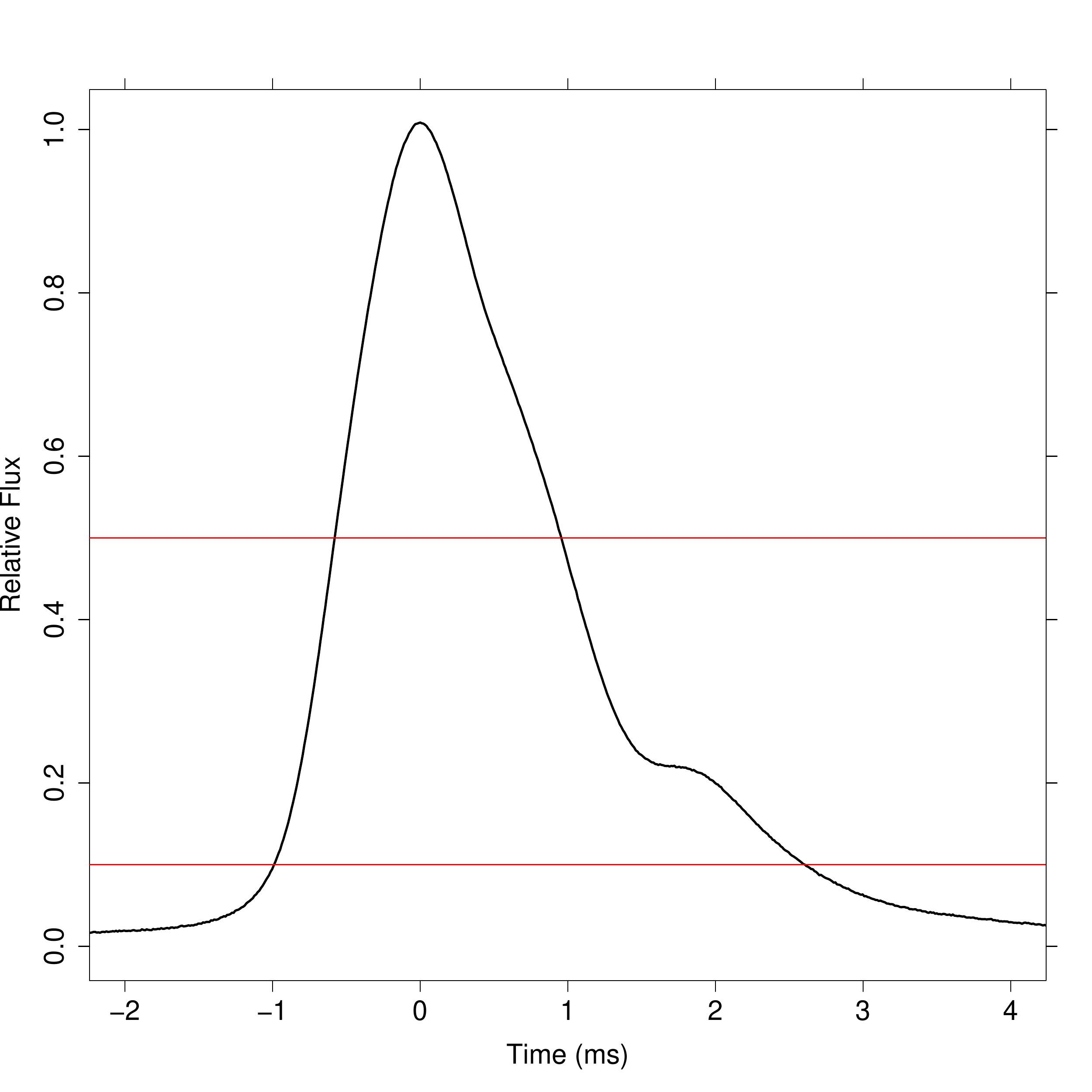}
 \llap{\parbox[b]{50.5mm}{$a$\\\rule{0ex}{63mm}}}
 \llap{\parbox[b]{32mm}{$b$\\\rule{0ex}{26mm}}}
 \llap{\parbox[b]{42mm}{$c$\\\rule{0ex}{48mm}}}
 \llap{\parbox[b]{75mm}{$d$\\\rule{0ex}{13mm}}}
 \caption{Typical daily integrated pulse profile (over 750000 pulses) of J0835$-$4510 at 1376~MHz showing the 10 and 50 per cent levels. The letters \textit{a,b,c,d} label emission zones discussed in \citet{krish1983}. See text.}
 \label{fig:profile}
 \end{figure}

There are three distinct visible and one non-visible features in the integrated profile of the Vela pulsar (see Figure~\ref{fig:profile}):
\begin{enumerate}
\renewcommand{\theenumi}{$\alph{enumi}$}
\item large main peak
\item `ledge' to the right
\item gentle point of inflection
\item bright pulse emission zone
\end{enumerate}

\citet{krish1983} discuss these four pulse components, their distributions, and their emission heights. 

The leading edge bright-pulse component ($d$) is not visible in the integrated profile because it only appears rarely. However, when the pulsar does emit a bright pulse, it is always on the leading edge, and it affects the pulse width strongly at both the 10 and 50 per cent levels. Therefore an increase in bright pulse rates should positively correlate with an increase in pulse width. It can be seen this is the case around the first small micro-glitch (MJD=56922), but it is clearly not the case with the large micro-glitch (MJD=57143) where the sudden decrease in pulse width did not appear alongside a sudden drop in bright pulse rate. A Lomb-Scargle periodogram of the bright pulse rates shows no significant periodicities. 

This is a paradox. The first micro-glitch coincides with a sudden increase in bright pulse rates, with no apparent change in pulse width. The second micro-glitch coincides with a sudden decrease in pulse width with no apparent change in bright pulse rate.

This can be explained in the model of \citet{wright2003} who considers that emission zones might be mathematically chaotic in nature, however this hypothesis relies on the emission zones occurring in the cone, whereas Vela (being a young pulsar) should have main core emission rather than conal emission \citep{rankin1990}.

Finally, note in Figure~\ref{fig:width}d the 50~day `quiet time' just before the first micro-glitch and the gradual overall `settling' of the bright pulse rate over time. This may settle to another `quiet time' which might indicate that another glitch is imminent.

\section{CONCLUSIONS}

We are carrying out an intensive single-pulse observation campaign of the Vela pulsar and have collected over 6000~hours of data. We have shown that the daily integrated pulse profile width changes both slowly over time and has a discontinuity after a micro-glitch. Bright-pulse rates are also shown to be affected by micro-glitches, but in an inconsistent manner.

We have found periodicities in the pulse width changes that match (within error bars) the exciting X-Ray results from \citet{durant2013} that imply free precession.

We hope that these results might shed some new light on the pulsar emission and glitching process, and to this end, we intend to produce further papers from this large data set.

\acknowledgments

We would like to thank Mr~Brett Reid, Mr~Eric Baynes, Dr~Jamie McCallum, Mrs Melissa Humphries, Dr Andrew Cole, and Dr~Lucia Plank of the Department of Physical Sciences at the University of Tasmania who have cheerfully assisted us in the collection and analysis of this data. Professor Joanna Rankin also provided stimulating input into this work. We would also like to acknowledge the Tasmanian Partnership for Advanced Computing (TPAC) at the University of Tasmania with funding from the Australian Government through its NCRIS and RDSI programs for the use of the 2.3~PB storage facility, without which this project would not be possible.

\bibliographystyle{apj}

\bibliography{/Users/jim/jim}

\end{document}